\documentclass[aps,prl,preprint,groupedaddress]{revtex4-1}

\usepackage{graphicx,color}

\newcommand{\m}[1]
{\mathrm{#1}}

\begin{document}

%Title of paper
\title{A single quantum dot as an optical thermometer for mK temperatures}

\author{Florian Haupt}
\author{Atac Imamoglu}
\author{Martin Kroner}
\affiliation{Institute of Quantum Electronics, ETH Z\"{u}rich,
CH-8093 Zurich, Switzerland}

\date{\today}

\begin{abstract}
Resonant laser spectroscopy of a negatively charged self-assembled
quantum dot is utilized to measure the temperature of a three
dimensional fermionic reservoir down to 100mK. With a magnetic field
applied to the quantum dot the single charged ground state is split
by the Zeeman energy. As the quantum dot is in tunnel contact with a
thermal electron reservoir, a thermal occupation of the quantum dot
spin states is enforced by co-tunneling processes. Resonant laser
induced fluorescence is used in order to measure the thermal quantum
dot spin state population.
\end{abstract}

\maketitle

% General introduction
%%%%%%%%%%%%%%%%%%%%%%%%%%%%%%%%%%%%%%%%%%
A recent development in quantum dot (QD) optics is that it is not
the QD itself that is subject to research but rather the interaction
with its environment \cite{Schulhauser2000, Koenraad2010, Latta2011,
Haupt2013}. The atom-like optical properties and well understood
electrical structure makes the QD an ideal probe for exploring the
rich physics of higher dimensional fermionic systems. In particular,
when it comes to highly correlated electron states formed by a QD
and a two-dimensional electron system such as the Kondo effect
(\cite{GoldhaberGordon1998, Cronenwett1998, Latta2011}), the
temperature of the electron system plays a major role and low
temperatures on the mK scale, that can only be reached in a dilution
refrigerator, are desirable. While the electronic energy scale in a
Fermi reservoir is determined by the temperature, the optical
transitions of an isolated QD is hardly influenced by the the
temperature of the itinerant electrons or that of the lattice. The
tight confinement ($\approx 30\m{meV} $) for the electron and hole
wave function in the QD, which is much larger than the thermal
energy at liquid Helium temperatures ($360\mu\m{eV}$), leads to
negligible thermal excitation. Furthermore, the QD-phonon
interaction only becomes relevant for temperatures larger than
$15\m{K}$, as the relevant phonon modes with sizable coupling
strength become occupied \cite{Borri2001, KronerTemp2009}. Hence,
the resonance line width of excitonic transitions in the QD is in
principle limited only by the spontaneous emission rate
\cite{Kuhlmann2013}. This stands in stark contrast to transport
spectroscopy of electrostatically defined QDs in a GaAs/AlGaAs
two-dimensional electron system: there, the width of Coulomb
blockade oscillations is governed by the Fermi-Dirac distribution of
electronic energies in the leads. Hence, transport spectroscopy
provides a highly sensitive and direct way to measure the
temperature of a fermionic reservoir in the mK range
\cite{Pekola1994}. However, transport spectroscopy is not always
applicable as a thermometer, in particular for optical experiments.

% Preview
%%%%%%%%%%%%%%%%%%%%%%%%%%%%%%%%%%%%%%%%%%
In this letter, we present an all optical approach to measure the
temperature of a fermionic reservoir that is tunnel coupled to a QD.
We perform resonant laser absorption spectroscopy on the
single-charged exciton ground state transitions of a single QD. In
this way, we measure the thermal occupation of the QD ground state
and obtain the temperature of the fermionic reservoir. We measure
reproducibly a temperature of $T\approx 130\m{mK}$. Furthermore,
this technique allows to study the heating induced by laser fields.

% Introduction to the experiment
%%%%%%%%%%%%%%%%%%%%%%%%%%%%%%%%%%%%%%%%%%

\begin{figure}
\includegraphics{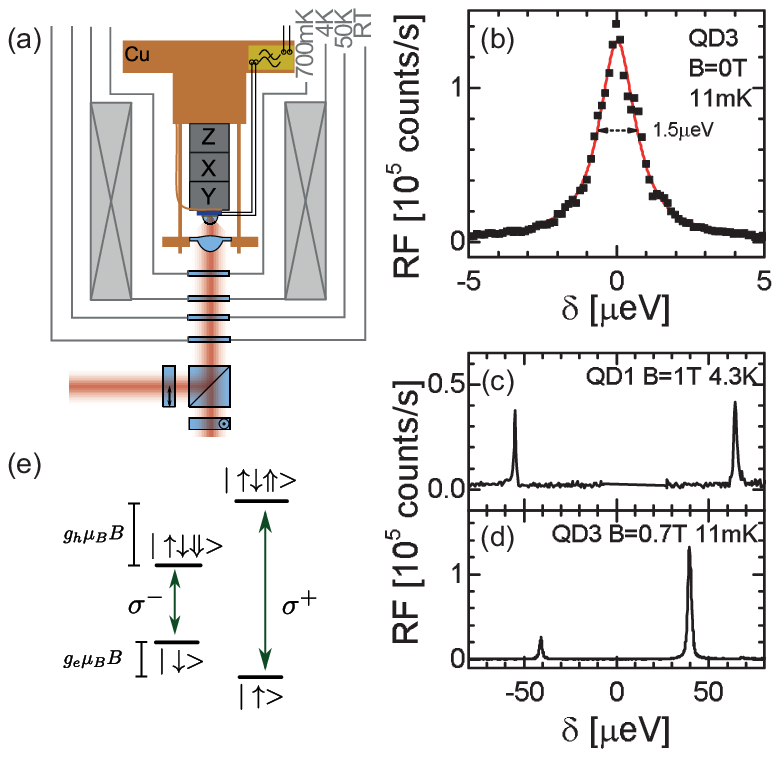}
\caption{\label{Fig1} (a) Schematic of the free-space optical
dilution refrigerator. The sample is mounted on xyz-piezo positioner
and is in thermal contact with the Cu base plate via thermal braids
and the electrical connections. The radiation shields with are
labeled by their temperature. The vector magnet is mounted on the 4K
stage. Only the room temperature (RT) window is vacuum sealed. The
optical excitation and collection are performed via a beam splitter.
Two orthogonal polarizing filter are used for the RF experiment as
described in the text. (b) Typical RF spectrum as a function of the
laser detuning at mK temperature with no magnetic field applied. (c)
and (d) RF spectrum  vs. laser detuning at different magnetic fields
and temperatures as labeled in the figures. (e) Four level scheme of
the negatively charged QD with the relevant optical selection rules
indicated.}
\end{figure}

The QD is embedded in a Schottky diode structure which allows for a
controlled charging of the QD with individual electrons
\cite{Drexler1994, Warburton2000, Hogele2004}. To this end the QD
layer is separated by a 35nm thick tunnel barrier of intrinsic GaAs
from a highly n-doped GaAs layer which acts as an electron
reservoir. On top of the sample a semitransparent metallic top gate
is deposited in order to apply a voltage with respect to the
electron reservoir. In order to obtain the optical spectrum of the
QD we scan a narrow band laser across the exciton transition of a
single charged QD and record the resonance fluorescence (RF) in
reflection \cite{Vamivakas2009, Fallahi2010}. Cross-polarized
excitation and detection is used in order to suppress the laser
background (Fig. \ref{Fig1} (a)). The collection efficiency is
enhanced by a half-hemispherical $\m{ZrO}_2$ solid immersion lens on
top of the sample and a distributed Bragg reflector (28 layers) that
has been grown below the QD layer (see \cite{Weibo2012} for details
on the sample structure). A typical spectrum, measured on a
negatively charged QD with no magnetic field applied, is shown if
Fig. \ref{Fig1} (b). We routinely observe a resonance line width
$\gamma\approx 1.5\mu e\m{V}$.

% Setup
%%%%%%%%%%%%%%%%%%%%%%%%%%%%%%%%%%%%%%%%%%
The sample is mounted in a cryogen-free dilution refrigerator
(Bluefors LD250) with a base temperature of 7mK. For free-space
optical access to the sample, the cryostat is equipped with optical
windows (fused silica) in all radiation shields of the different
thermal stages (Fig. \ref{Fig1} (a)). For optimal transmission of
the laser and RF photons the windows are anti-reflection coated in
the wavelength range $\lambda = 600\m{nm} - 1000\m{nm}$. In order to
suppress etaloning, the windows are wedged by $0.5^\circ$. With the
windows installed the base temperature of the cryostat increases to
11mK, due to the additional heat load, which is measured to be
$\approx 1\mu W$. A single aspheric lens objective ($NA=0.68$) is
used to focus the laser onto the sample surface and to collect the
back scattered photons that are ported out of the cryostat to be
analyzed at room temperature. The sample is mounted on a piezo
driven positioner stage (Attocube ANPx101/res (2x), ANPz101/res
(1x)) with a resistive position read out. The resistive readout
however induces significant heating and is switched off in order to
reach the base temperature. To thermalize the sample, we connect it
to the base plate via thermal braids (Attocube ATC100). The signal
cables for electrical contacting the sample are, for each line
individually, thermalized by a second order low pass filter towards
the grounded base plate. The slip-stick mode of operation of the
positioners requires connector cables with a large frequency
bandwidth. Due to the significant capacitance of the piezo actuators
($C\approx 100\m{nF}$) low Ohmic wiring, ideally below $2\Omega$ is
necessary and hence low pass filtering as for the signal cables is
not applicable. Superconductive wiring is used, in order to minimize
the heat load and to guarantee the functionality of the positioners
at low temperatures. The cryostat is further equipped with a 9T/3T
vector magnet.

% Temperature measurement scheme
%%%%%%%%%%%%%%%%%%%%%%%%%%%%%%%%%%%%%%%%%%
In order to measure the electron temperature of the back contact we
apply a magnetic field parallel to the optical excitation direction
(Faraday geometry) which leads to a Zeeman splitting of the single
electron ground state of the QD \cite{Bayer2002, Hogele2004}. Since
the QD is tunnel coupled to the back contact the occupation of the
two electron spin ground states will correspond to that of a thermal
state. We measure this ground state population by performing
absorption spectroscopy on the fundamental charged exciton
transitions. The wavelength of the laser field determines which QD
transition is driven. The projection of the laser polarization onto
the polarization vector representing the optical selection rules of
the selected QD transition scales the strength of the laser drive
and in turn the strength of the absorption signal. The signal
strength is further determined by the population of the
corresponding ground state. Hence, keeping the laser polarization
constant, a change in the absorption signal signifies a change of
the ground state population \cite{Hoegele2005}. For small magnetic
fields and high temperatures the two electron spin states have equal
population and consequently the two resonances will exhibit equal
strength (Fig. \ref{Fig1} (c)). In order to ensure the polarization
independence of the measurement we choose the laser field to be
linearly polarized, while the optical selection rules for the QD
exciton transitions are circularly polarized (Fig. \ref{Fig1} (e)).
Further, the laser Rabi frequency $\Omega$ is small compared to the
radiative emission rate of the QD exciton. This corresponds to
driving the exciton transition below saturation in the so-called
linear regime where the absorption signal depends linearly on the
laser power (or the square of the Rabi frequency)
\cite{KronerSat2008}. If the Zeeman splitting of the electron ground
state exceeds the temperature, the population of the two electron
spin states ($\rho_\downarrow,\rho_\uparrow$) will differ, resulting
in different absorption signals as shown in Fig. \ref{Fig1} (d).
There, the temperature of the cryostat is $T_\m{cryo}=11\m{mK}$ and
the magnetic field is $B=0.7\m{T}$. The thermal population
$\rho_{\downarrow / \uparrow}$ is given by the Fermi-Dirac
distribution and the relative amplitudes of the absorption
resonances $A_\m{blue}$ for the high energy transition
(corresponding to the spin up ground state) and $A_\m{red}$ for the
low energy transition (corresponding to the spin down ground state)
can be written as:

\begin{equation}
\label{pop} \rho_{\downarrow /
\uparrow}=\frac{A_{\m{red}/\m{blue}}}{A_{\m{red}}+A_{\m{blue}}}=\frac{1}{\exp\left(
 \pm \frac{g_\m{e} \mu_\m{B} B}{k_\m{B} T}\right)+1}.
\end{equation}

Here $g_\m{e}$ is the electron g-factor and $\mu_\m{B}$ and
$k_\m{B}$ are the Bohr magneton and the Boltzmann constant,
respectively. The externally applied magnetic field is given by $B$
and the temperature of the Fermi sea of electrons in the back
contact is $T$.

\begin{figure}
\includegraphics{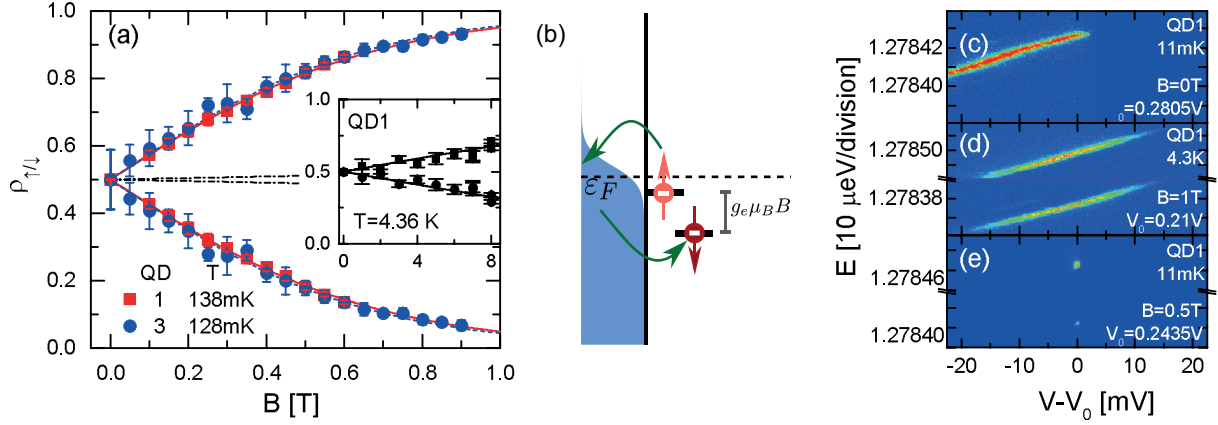}
\caption{\label{Fig2} (a) Relative absorption amplitude of the two
optical transitions (as shown in Fig. \ref{Fig1} (c) and (d)) as
function of the magnetic field for two different QDs (red squares
and blue dots). The correspondingly colored lines indicate the
thermal occupation of the QD electronic spin states based on Eq.
\ref{pop}. The inset shows a similar measurement obtained only with
the pulse tube cooling (also indicated by the dash-dotted line in
the main figure). (b) Schematic to illustrate the co-tunneling
process. An electron from the QD (red arrow) can tunnel into the
Fermi reservoir (occupation relative to the Fermi energy
$\epsilon_\m{F}$ indicated by the blue area) and be replaced by
another electron of opposite spin. The probability of the resulting
spin state occupation scales with the Zeeman splitting between the
two states. (c)-(e) Color coded RF counts (blue: low counts, red:
high counts) as function of laser energy and gate voltage at the
edge of the $X^-$-plateau. All measurements where taken on QD1. The
magnetic field and temperature are labeled, so is the gate voltage
offset $V_0$ indicating the charging voltage for the second
electron. This voltage varies over time and hence is different for
these data sets that where taken over the course of several weeks.}
\end{figure}

% Temperature measurement results
%%%%%%%%%%%%%%%%%%%%%%%%%%%%%%%%%%%%%%%%%%
In Fig. \ref{Fig2} (a), $\rho_{\downarrow / \uparrow}$ is plotted
for two different QDs (QD1, QD3) as a function of the magnetic
field. By fitting Eq. \ref{pop} to the data points we obtain the
temperatures $T$ as labeled in the plot. The electronic g-factors of
the QDs ($g_\m{e,QD1}=0.61$,$g_\m{e,QD3}=0.59$) were determined by a
two-laser re-pump experiment as described in \cite{KronerESR2008}.
As a comparison the inset in Fig. \ref{Fig2} (a) shows
$\rho_{\downarrow / \uparrow}$, when the sample is cooled only by
the pulse tube cooler ($T_\m{cryo}\approx 4K$). Even at magnetic
fields up to 8T the amplitude ratio is not changing appreciably at
these temperatures and hence the accuracy of this measurement is
limited. For a base temperature of $T_\m{cryo}=11\m{mK}$ we find the
electron temperature of the sample to be $T\approx (130\pm 7)\m{mK}$
with a very high accuracy of $5\%$. However, the measured electron
temperature of the samples is more than one order of magnitude
higher than the base temperature of the cryostat.

% Influences of cotunneling and spin pumping
%%%%%%%%%%%%%%%%%%%%%%%%%%%%%%%%%%%%%%%%%%
Before discussing the implications of this unexpectedly high
measured temperature, we explain the framework of the temperature
measurement where several important points need to be considered for
an accurate temperature estimate. As discussed before, the thermal
occupation of the QD electronic states is imposed by the coupling to
the thermal electron system of the back contact. In particular, a
fast and efficient thermalization of the QD electron via
co-tunneling processes to the electron reservoir has to be
guaranteed. At finite magnetic fields in the center of the voltage
range of the stable single electron state in the QD (the
$X^-$-plateau) the absorption signal vanishes due to optical spin
pumping and slow spin relaxation. This is a consequence of the fact
that both, exchange with the Fermi sea and hyperfine flip-flop terms
are suppressed \cite{Atature2006, Dreiser2008}. However, if the QD
electronic state is tuned close to the Fermi level fast co-tunneling
\cite{Smith2004} will lead to a thermal occupation of states in the
QD and the absorption signal is restored. This is sketched in Fig.
\ref{Fig2} (b). In order to ensure a thermal state in the QD we have
to apply a gate voltage such that the measured resonance signal (of
the two transitions combined) becomes similar to the signal measured
with no magnetic field applied. In Fig. \ref{Fig2} (c)-(e) the RF
counts are color coded as a function of laser energy and gate
voltage for several temperatures and magnetic fields. In panel (c)
the end of the charge stability plateau of the singly charged QD is
shown at $B=0\m{T}$ and $T_\m{cryo}=11\m{mK}$. The absorption
amplitude is constant over the $X^-$-plateau and then drops off when
the QD ground state is doubly charged at more positive gate
voltages. This abrupt drop of the RF-signal resembles the sharp
Fermi-Dirac distribution of the fermionic reservoir of electrons,
which is the source for the second electron tunneling into the QD at
this voltage. Here, the distribution of occupied states in the back
contact is probed by the electron state of the QD whose energy
relative to the Fermi energy is given by the electrostatic energy,
$E_\m{el}=e V_\m{g} / \eta$ with the gate voltage $V_\m{g}$ and the
lever arm $\eta=5$ \cite{Seidl2005}. The energy resolution is given
by the voltage fluctuations in the sample to be on the order of
$\approx 100\mu e \m{V}$ restricting the lowest measurable
temperature to be measured this way to be $\approx 1\m{K}$.
\cite{ElectrostaticsVsQCSE}. In Fig. \ref{Fig2} (d) the co-tunneling
regime is plotted at $B=1\m{T}$ and $T_\m{cryo}=4.3\m{K}$. The
rather smooth decay of the signal strength towards the edges of the
co-tunneling regime is due to the large temperature as discussed
before. Also, the amplitudes of the two resonances are almost equal
at this low magnetic field. At base temperature however, as shown in
Fig. \ref{Fig2} (e), the effect of the temperature is striking. For
$B=0.5\m{T}$ and $T_\m{cryo}=11\m{mK}$ the voltage range where we
find the QD ground state in a thermal state is reduced to one
optical line width (limited by the spectral fluctuations) and the
blue, high energy transition clearly shows a larger absorption
amplitude than the red, low energy transition. In order to obtain a
precise reading of the temperature we perform a measurement as shown
in Fig. \ref{Fig2} (d) or (e) for each magnetic field. We
post-select the spectra in the center of the co-tunneling range
leading to a mean value and a standard deviation (data points and
error bars in Fig. \ref{Fig2} (a)) for $\rho_{\downarrow /
\uparrow}$.

% Influences of dragging
%%%%%%%%%%%%%%%%%%%%%%%%%%%%%%%%%%%%%%%%%%
An important prerequisite for the described analysis is that the
resonance line shape is not distorted by other influences such as
the nuclear spin environment via the dragging effect
\cite{Latta2009, Hoegele2012}. In order to circumvent dragging of
the QD resonance we scan the laser fast across the resonance so that
the rather slow nuclear spin dynamics cannot follow
\cite{Hoegele2012}. More precisely, we scan the laser energy with a
rate of $\approx 100 \mu e \m{V} / \m{s}$ and average over 10 to 20
scans to obtain one spectrum.

\begin{figure}
\includegraphics{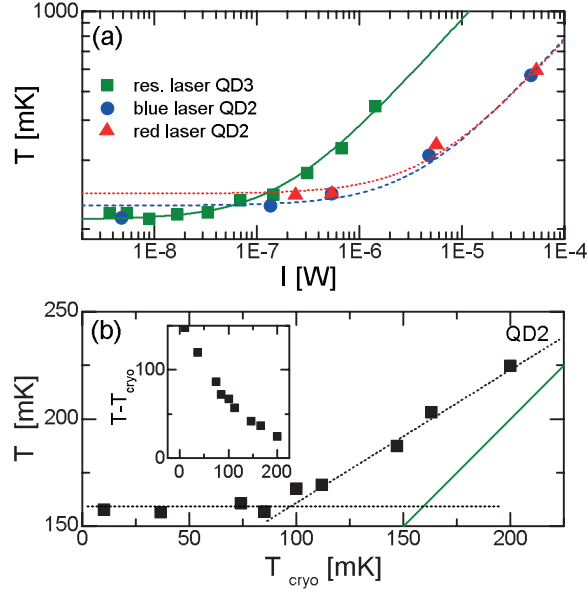}
\caption{\label{Fig3} (a) Measured temperature as a function of the
incident laser power. The green squares stand for the experiment
with increasing resonant probe laser power measured on QD3. The red
and blue data point where acquired for a constant probe laser power
($P=100\m{nW}$) for different red ($\lambda_\m{red}=980\m{nm}$) or
blue ($\lambda_\m{blue}=904\m{nm}$) detuned heating lasers on QD2.
The lines are guide to the eyes. (b)  Measured temperature $T$ as a
function of the mixing chamber temperature $T_\m{cryo}$ (black
squares). The dotted lines are guides to the eye, the green line
indicates $T_\m{cryo}$. The inset shows $T-T_\m{cryo}$ vs.
$T_\m{cryo}$. This data set was obtained on QD2.}
\end{figure}

% Laser power and local heating
%%%%%%%%%%%%%%%%%%%%%%%%%%%%%%%%%%%%%%%%%%
From a four level master equation simulation including the
co-tunneling mechanism we conclude that the influence of the
resonant laser on the ground state population is negligible for
realistic parameters. To experimentally verify that fast and
efficient spin pumping does not alter the temperature measurement we
also performed a measurement in Voigt rather than in Faraday
geometry (data not shown). With an applied in-plane magnetic field
the spin pumping rate is as fast as the radiative decay rate
\cite{Xu2007} and about a factor of 200 faster than in Faraday
geometry. Still the temperature measurement resulted in similar
values as before.\\
However, the laser power can certainly lead to local heating of the
sample, in particular, due to a considerable absorption of the laser
light by the metallic top gate. To exclude laser heating we measured
the sample temperature for different laser powers on QD3. In Fig.
\ref{Fig3} (a) the obtained temperature values are plotted as
function of the laser power measured outside the cryostat. We find
that for laser powers below $50\m{nW}$ the measured temperature
remains constant at a value of $T=130\m{mK}$. Hence, we can exclude
that the laser power is restricting the temperature measurement. For
larger laser power, we do indeed observe an increase of the
temperature hinting towards laser heating effects. In order to
exclude distortion of the temperature measurement by saturation of
the exciton transition \cite{KronerSat2008} we performed a control
experiment with a second, off-resonant laser. (The two laser
experiment was performed on QD2.) While the resonant probe laser
remains at a power of $100 \m{nW}$ either a red
($\lambda_\m{red}=980\m{nm}$) or blue ($\lambda_\m{red}=904\m{nm}$)
detuned laser was varied in power. The resulting temperature as a
function of the off-resonant laser power is shown in Fig. \ref{Fig3}
(a) by the red and blue points respectively. This procedure allowed
us to measure the local laser induced heating and resulting
temperature up to a laser power of $50\mu \m{W}$ a power which is
two orders of magnitude above the saturation power and also larger
than the cooling power of the dilution unit ($\sim8\mu \m{W}$) at
base temperature. We observe that the heating due to the resonant
laser is considerably larger than due to the off-resonant laser,
while the blue and red detuned laser have a very similar heating
effect. This is remarkable since the blue detuned laser is very
close to the excited QD states. Hence, we would have expected
stronger heating effect as for the red detuned laser which is
further away from most residual absorption transitions. Both lasers
however lie beyond the bandwidth of the Bragg-mirror in the sample.
This means that the laser light will be transmitted through the
sample and finally be absorbed on the sample holder which is very
well thermalized. The resonant laser however will be reflected by
the mirror and pass through the gate a second time where it can be
absorbed again. This does not explain though the large difference
between the resonant and the off-resonant laser heating effects
which stems most probably from saturation effects. From the similar
results for the two off-resonant lasers we can conclude that the
most important mechanism for laser induced heating in these cases is
the absorption by the gate electrode.

% Temperature dependence and temperature gradient
%%%%%%%%%%%%%%%%%%%%%%%%%%%%%%%%%%%%%%%%%%
The measurements at low laser power indicate that the electron
temperature is not limited by the measurement itself but is rater
limited by the imperfect thermal coupling of the sample to the base
plate. It has to be said that since the sample has to be movable and
hence can not be mounted in close proximity of the cold plate it is
not too surprising that a thermal gradient forms across the
superstructure of the microscope.  The cooling power mediated by the
cables, the most significant cooling mechanism for the electron
system in the sample at low temperatures, is limited, despite the
careful thermalization and electric filtering. To explore these
limits we measured the sample temperature for different base
temperatures on QD2. The data is shown in Fig. \ref{Fig3} (b). In
order to ensure the independence of the individual measurements of
this experiment the data was taken in random order. For temperatures
of the base plate below 100mK the measured sample temperature $T$
remains constant at 160mK \cite{HighTemp}. Above 100mK the sample
temperature starts to increase and approaches the base temperature
(indicated by the green line in Fig. \ref{Fig3} (b)). For clarity
the inset shows the difference between the sample and the base plate
temperature. Since the cooling power of the dilution unit depends
strongly on temperature (quadratically at these low temperatures)
for increasing temperature and hence increasing cooling power the
gradient that forms across the thermal anchoring of the sample,
diminishes from 150mK at $T_\m{cryo}=11\m{mK}$ down to 25mK at
$T_\m{mix}=200\m{mK}$.

\begin{figure}
\includegraphics{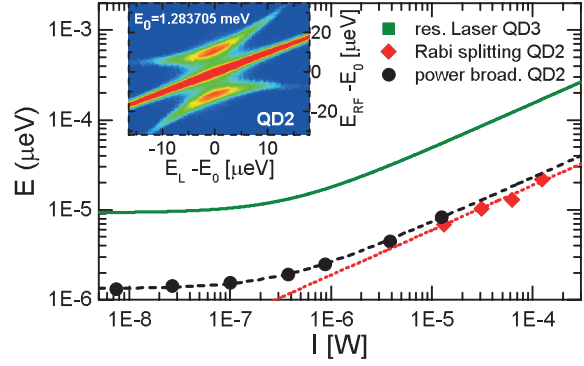}
\caption{\label{Fig4} Power broadening (black dots) of a resonantly
driven QD transition and Rabi energy (red diamonds) of the resonant
laser driving a QD vs. the corresponding laser power. The dotted and
dashed lines are guides to the eye. The green line represents the
thermal energy measured as function of laser power as shown in Fig.
\ref{Fig3} (b). The inset shows the color coded QD emission as
function of the photon energy $E_\m{RF}$ relative to the QD
resonance energy  $E_0$ and the laser detuning from the resonance
($E_\m{L}-E_0$). The data was obtained using a scanning Fabry-Perot
spectral filter. The splitting at zero laser detuning gives the
laser Rabi energy.}
\end{figure}

% Implications
%%%%%%%%%%%%%%%%%%%%%%%%%%%%%%%%%%%%%%%%%%
The setup described in this letter allows for free-space laser
spectroscopy on a single QD at mK temperatures. In particular the
free space optical access allows for a high degree of polarization
control and in turn for measurement of the resonance fluorescence of
a QD. This allows us not only to perform laser spectroscopy but also
to measure the power spectrum of the resonantly scattered photons.
It has been proposed that a single QD strongly tunnel-coupled to an
electron reservoir, that exhibits Kondo-correlations for weak
optical excitations \cite{Latta2011}, exhibits a non-equilibrium
quantum correlated state in the strong optical driving limit
\cite{Sbierski2013}. Such a state is expected to alter the power
spectrum from the well known Mollow-triplet \cite{Mollow1969,
Muller2008} to a two peak structure with a characteristic power
dependence. In order to explore the feasibility to reach a sample
temperature lower than the energy scale of the Rabi energy as well
as the Kondo temperature, we measured the Rabi energy of the laser
from the power broadening as well as directly from the
Mollow-triplet. In the inset of Fig. \ref{Fig4} a typical energy
resolved resonance fluorescence measurement is shown as a function
of the laser detuning of QD2. This way we determined the Rabi energy
$\hbar \Omega$ for a strongly driven QD transition as a function of
laser power (red data points in Fig. \ref{Fig4}). The black data
points in Fig. \ref{Fig4} indicate the extracted linewidth from a
resonant saturation experiment signifying power broadening
\cite{KronerSat2008}. In order to compare the energy scale of the
Rabi energy with the temperature, we also plot the thermal energy
corresponding to the laser power dependent temperature from Fig.
\ref{Fig3} (a). From this comparison it is obvious that for any
laser power the Rabi energy is roughly one order of magnitude
smaller than the thermal energy. In order to reach a situation where
the Rabi energy can be comparable to the thermal energy of the
electron reservoir, improvements need to be made to reduce the
electron temperature and the laser induced heating effects at high
laser powers. To address the latter, the Schottky structure with the
metallic top gate can be replaced by a pin-diode structure. Also,
the light matter interaction can be increased by replacing the
$\m{ZrO}_2$ SIL with one made from GaAs \cite{Vamivakas2007} or
using spacial confinement of the light mode \cite{Bakker2014}, which
reduces the required laser power to obtain a certain Rabi energy. In
order to further reduce the electron temperature the sample
structure can be optimized for a lower resistance electrical
connection to the Fermi-reservoir.

% Implications
%%%%%%%%%%%%%%%%%%%%%%%%%%%%%%%%%%%%%%%%%%
In conclusion we demonstrated an all optical scheme to measure the
electron temperature in a semiconductor down to the mK range. The
presented results open new ways for exploring the physics of
semiconductor structures at ultra low temperatures by high
resolution resonant optical spectroscopy. In particular the ability
to measure the resonance fluorescence gives direct access to the
power spectrum of a single QD at mK temperatures in a cryogen free
dilution refrigerator.

% Thanks
%%%%%%%%%%%%%%%%%%%%%%%%%%%%%%%%%%%%%%%%%%
We thank Clemens R\"ossler, Khaled Karrai and the Bluefors team for
discussions and technical support.

\newpage

\begin{thebibliography}{}
\bibitem{ElectrostaticsVsQCSE} This energy scale is not to be confused with the optical
resolution of the exciton transition where the sensitivity to the
electric field is given by the quantum confined Stark effect.
\bibitem{HighTemp} The slightly larger temperature compared to other measurements presented here arise from a larger laser power that was used in this experiment.
\end{thebibliography}
\end{document}